\documentclass{iaus238}

\usepackage{graphicx}

\title[The Galactic Center]{The Galactic Center}

\author[R.~Genzel \& V. Karas]{Reinhard 
Genzel$^{1,2}$ \and Vladim\'{\i}r Karas$^3$}

\affiliation{$^1$~Max-Planck Institut f\"ur Extraterrestrische Physik, Garching, Germany\\[\affilskip]
$^2$~Department of Physics, University of California, Berkeley, USA\\[\affilskip] 
$^3$~Astronomical Institute, Academy of Sciences, Prague, Czech Republic}

\pubyear{2006}
\volume{238}
\jname{Black Holes: from Stars to Galaxies -- across the Range of Masses}
\editors{V. Karas \& G. Matt, eds.}
\begin{document}

\maketitle

\begin{abstract}
In the past decade high resolution measurements in the infrared
employing adaptive optics imaging on $10$m telescopes have allowed
determining the three dimensional orbits stars within ten light hours of
the compact radio source at the center of the Milky Way. These
observations show the presence of a three million solar mass black hole
in Sagittarius~A* beyond any reasonable doubt. The Galactic Center thus
constitutes the best astrophysical evidence for the existence of black
holes which have long been postulated, and is also an ideal `lab' for
studying the physics in the vicinity of such an object. Remarkably,
young massive stars are present there and probably have formed in the
innermost stellar cusp. Variable infrared and X-ray emission from 
Sagittarius~A* are a new probe of the physical processes and space-time 
curvature just outside the event horizon.
\keywords{Galaxy: center -- black hole physics}
\end{abstract}

\firstsection 
\section{Introduction -- Sagittarius A*}
The central light years of our Galaxy contain a dense and luminous star
cluster, as well as several components of neutral, ionized and extremely
hot gas (Genzel, Hollenbach \& Townes \cite{gen94}). The Galactic Center
also contains a very compact radio source, Sagittarius~A* (Sgr~A*;
Balick \& Brown \cite{bal74}) which is located at the center of the
nuclear star cluster and ionized gas environment. Short-wavelength
centimeter and millimeter VLBI observations have established that its
intrinsic radio size is a mere $10$~light minutes (Bower {\etal}
\cite{bow04}; Shen {\etal} \cite{she05}). Sgr~A* is also an X-ray
emission source, albeit of only modest luminosity (Baganoff {\etal}
\cite{bag01}). Most recently, Aharonian {\etal} (\cite{aha04}) have
discovered a source of TeV $\gamma$-ray emission within $10$~arcsec of
Sgr~A*. It is not yet clear whether these most energetic $\gamma$-rays
come from Sgr~A* itself or whether they are associated with the nearby
supernova remnant, Sgr~A East.

Sgr~A* thus may be a supermassive black hole analogous to QSOs, albeit
of much lower mass and luminosity. Because of its proximity -- the
distance to the Galactic Center is about $10^5$ times closer than the
nearest quasars -- high resolution observations of the Milky Way nucleus
offer the unique opportunity of stringently testing the black hole
paradigm and of studying stars and gas in the immediate vicinity of a
black hole, at a level of detail that will not be accessible in any
other galactic nucleus in the foreseeable future. 

Since the center of the Milky Way is highly obscured by interstellar
dust particles in the plane of the Galactic disk, observations in the
visible light are not possible. Investigations require measurements at
longer wavelengths -- the infrared and microwave bands, or at shorter
wavelengths -- hard X-rays and $\gamma$-rays, where the veil of dust is
transparent. The dramatic progress in our knowledge of the Galactic
Center over the past two decades is a direct consequence of the
development of novel facilities, instruments and techniques across the
whole range of the electromagnetic spectrum.

\begin{figure}
\includegraphics[height=2.7in]{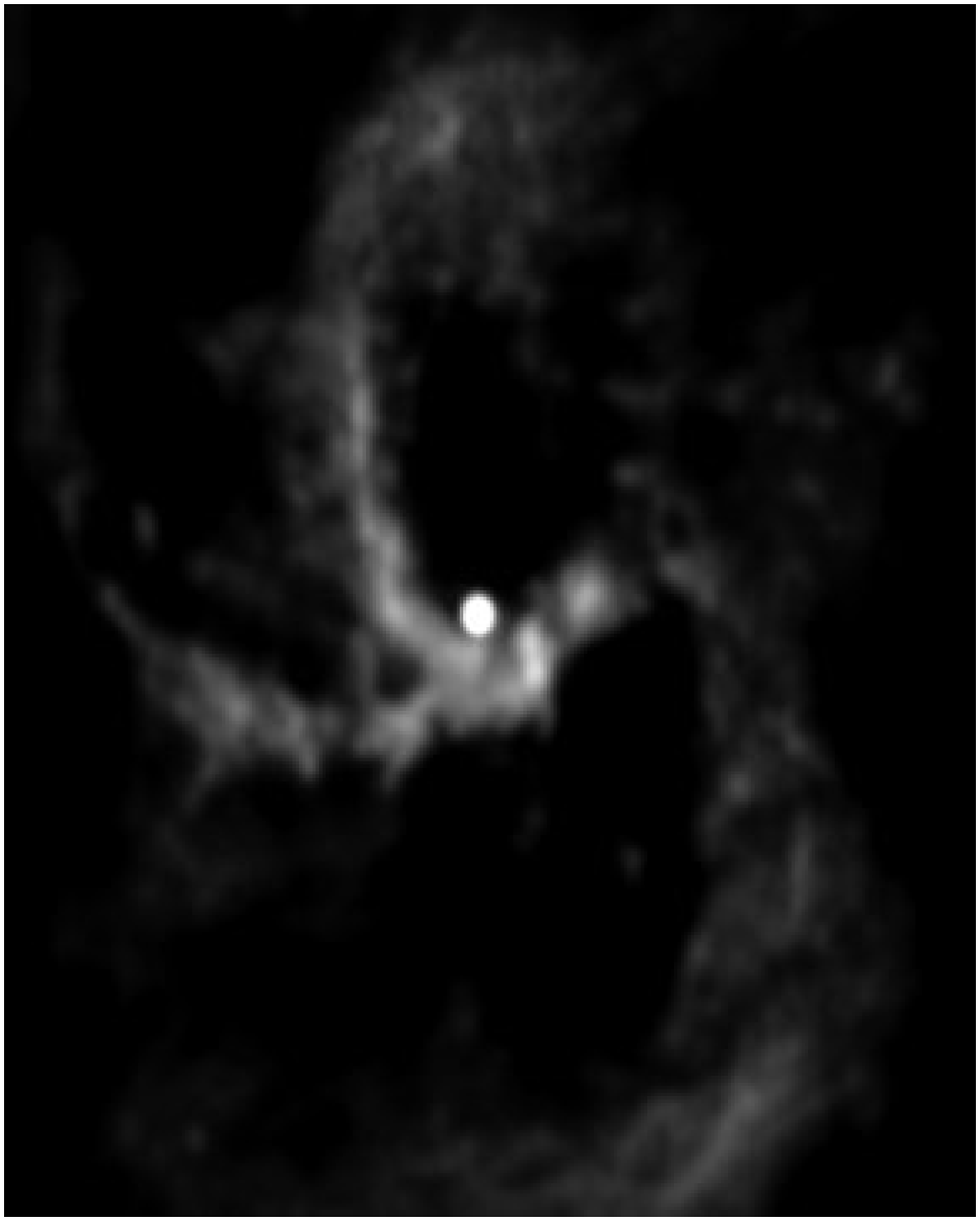}
\hfill
\includegraphics[height=2.7in]{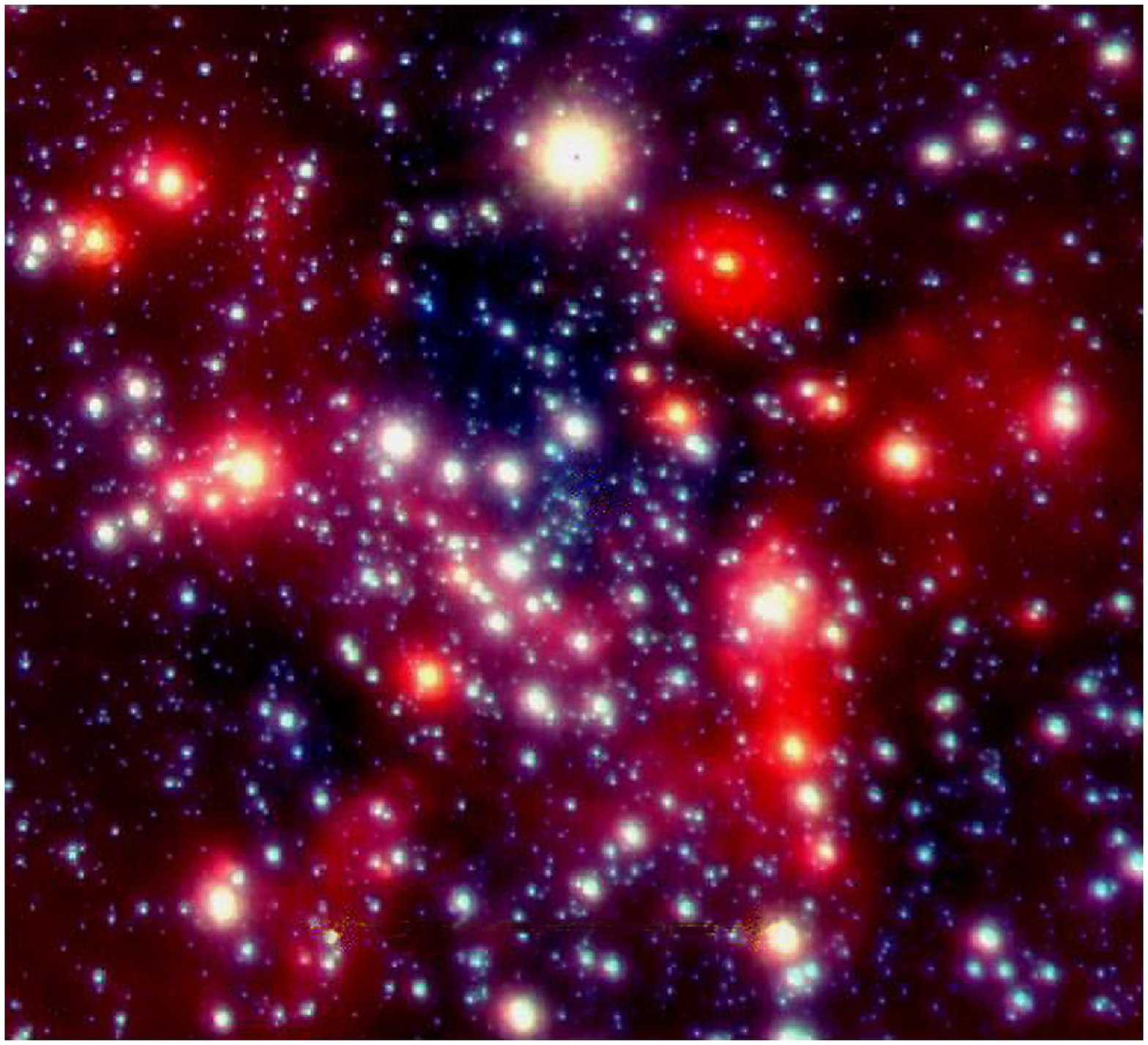}
\caption{Left: VLA radio continuum map of the central parsec (Roberts \&
Goss \cite{rob93}). The radio emission delineates ionized gaseous
streams orbiting the compact radio source Sgr~A*. Spectroscopic
measurements in the radio band (Wollman {\etal} \cite{wol77}) provided
the first dynamical evidence from large gas velocities that there might
be a hidden mass of $3$--$4$ million solar masses located near Sgr~A*.
Right: A diffraction limited image of Sgr~A* ($\sim0.05$~arcsec
resolution) from the $8$m ESO VLT, taken with the NACO AO-camera and an
infrared wavefront sensor at $1.6$/$2.2$/$3.7$~${\mu}m$ (Genzel {\etal}
\cite{gen03b}).  The central black hole is located in the centre of the
box. NACO is a collaboration between ONERA (Paris), Observatoire de
Paris, Observatoire Grenoble, MPE (Garching), and MPIA (Heidelberg)
(Lenzen {\etal} 1998; Rousset {\etal}
\cite{rou98}).}\label{fig1}
\end{figure}

\section{High angular resolution astronomy}
The key to the nature of Sgr~A* obviously lies in very high angular
resolution measurements. The Schwarzschild radius of a $3.6$ million
solar mass black hole at the Galactic Center subtends a mere
$10^{-5}$~arcsec. For the high-resolution imaging from the ground it is
necessary to correct for the distortions of an incoming electromagnetic
wave by the refractive and dynamic Earth atmosphere. VLBI overcomes this
hurdle by phase-referencing to nearby QSOs; sub-milliarcsecond
resolution can now be routinely achieved. 

In the optical/near-infrared wavebands the atmosphere smears out
long-exposure images to a diameter at least ten times greater than the
diffraction limited resolution of large ground-based telescopes 
(Fig.~\ref{fig1}). From the early 1990s onward initially speckle imaging
(recording short exposure images, which are subsequently processed and
co-added to retrieve the diffraction limited resolution) and then later
adaptive optics (AO, correcting the wave distortions on-line) became
available. With these techniques it is possible to achieve diffraction
limited resolution on large ground-based telescopes. The diffraction
limited images are much sharper and also much deeper than the seeing
limited images. In the case of AO (Beckers \cite{bec93}) the incoming
wavefront of a bright star near the source of interest is analyzed, the
necessary corrections for undoing the aberrations of the atmosphere are
computed (on time scales shorter than the atmospheric coherence time of
a few milli-seconds) and these corrections are then applied to a
deformable optical element (e.g. a mirror) in the light path. 

The requirements on the brightness of the AO star and on the maximum
allowable separation between star and source are quite stringent,
resulting in a very small sky coverage of natural star AO. Fortunately,
in the Galactic Center there is a bright infrared star only $6$~arcsec
away from Sgr~A*, such that good AO correction can be achieved with an
infrared wavefront sensor system. Artificial laser beacons can overcome
the sky coverage problem to a considerable extent. For this purpose, a
laser beam is projected from the telescope into the upper atmosphere and
the backscattered laser light can then be used for AO correction. The
Keck telescope team has already begun successfully exploiting the new
laser guide star technique for Galactic Center research (Ghez {\etal}
\cite{ghe05a}).
After AO correction, the images are an order of magnitude sharper and
also much deeper than in conventional seeing limited measurements. The
combination of AO techniques with advanced imaging and spectroscopic
instruments (e.g. integral field imaging spectroscopy) have resulted in
a major breakthrough in high resolution studies of the Galactic Center. 

\section{Nuclear star cluster and the paradox of youth}
One of the big surprises is a fairly large number of bright stars in
Sgr~A*, a number of which were already apparent on the discovery
infrared images of Becklin \& Neugebauer (\cite{bec75}, \cite{bec78}).
High-resolution infrared spectroscopy reveals that many of these bright
stars are actually somewhat older, late-type supergiants and AGB stars.
Starting with the discovery of the AF-star (Allen {\etal} \cite{all90};
Forrest {\etal} \cite{for87}), however, an ever increasing number of the
bright stars have been identified as being young, massive and early
type. The most recent counts from the deep SINFONI integral-field
spectroscopy yields about one hundred OB stars, including various
luminous blue supergiants and Wolf-Rayet stars, but also normal
main-sequence OB stars (Paumard {\etal} \cite{pau06a}). The nuclear star
cluster is one of the richest concentrations of young massive stars in
the Milky Way.

The deep adaptive optics images also trace the surface density
distribution of the fainter stars, to about K~$17$--$18$~mag,
corresponding to late B or early A stars (masses of $3$--$6$ solar
masses), which are a better probe of the density distribution of the
overall mass density of the star cluster. While the surface brightness
distribution of the star cluster is not centered on Sgr~A*, the surface
density distribution is. There is clearly a cusp of stars centered on
the compact radio source (Genzel {\etal} \cite{gen03b}; Sch\"odel {\etal}
\cite{sch06}). The inferred volume density of the cusp is a power-law
${\propto}R^{-1.4\pm0.1}$, consistent with the expectation for a stellar
cusp around a massive black hole (Alexander \cite{ale05}). 

If there is indeed a central black hole associated with Sgr~A* the
presence of so many young stars in its immediate vicinity constitutes a
significant puzzle (Allen \& Sanders \cite{all86}; Morris \cite{mor93};
Alexander \cite{ale05}). For gravitational collapse to occur in the
presence of the tidal shear from the central mass, gas clouds have to be
denser than  $\sim10^9(R/(10^{\prime\prime}))^{-3}$ hydrogen atoms per
cm$^{-3}$.  This `Roche' limit exceeds the density of any gas currently
observed in the central region. Recent near-diffraction limited AO
spectroscopy with both the Keck and VLT shows that almost all of the
cusp stars brighter than $\mbox{K}\sim16$~mag appear to be normal, main
sequence B stars (Ghez {\etal} \cite{ghe03}; Eisenhauer {\etal}
\cite{eis05a}). If these stars formed in situ, the required cloud
densities approach the conditions in outer stellar atmospheres. 

Several scenarios have been proposed to account for this paradox of
youth. In spite of this effort the origin of central stars
(S-stars) is not well understood: models have difficulties in
reconciling different aspects of the Galaxy Centre -- on one side it is
a low level of present activity, indicating a very small accretion rate,
and on the other side it is the spectral classification that suggests
these stars have been formed relatively recently; see Alexander
(\cite{ale05}) for a detailed discussion and references. The most
prominent ideas to resolve the apparent problem are in situ formation in
a dense gas accretion disk that can overcome the tidal limits,
re-juvenation  of older stars by collisions or stripping, and rapid
in-spiral of a compact, massive star cluster that formed outside the
central region and various scattering a three body interaction
mechanisms, including resonant relaxation (Alexander \cite{ale05}).
Several other mechanisms have been proposed that could set stars on 
highly eccentric orbits and bring them to the neighbourhood of the
central black hole (e.g., Hansen \& Milosavljevi\'c
\cite{han03}; McMillan \& Portegies Zwart \cite{mcm03}; Alexander \&
Livio \cite{ale04}; \v{S}ubr \& Karas \cite{sub05}), but the problem  of
the S-stars remains open.

\begin{figure*}
\includegraphics[height=2.8in]{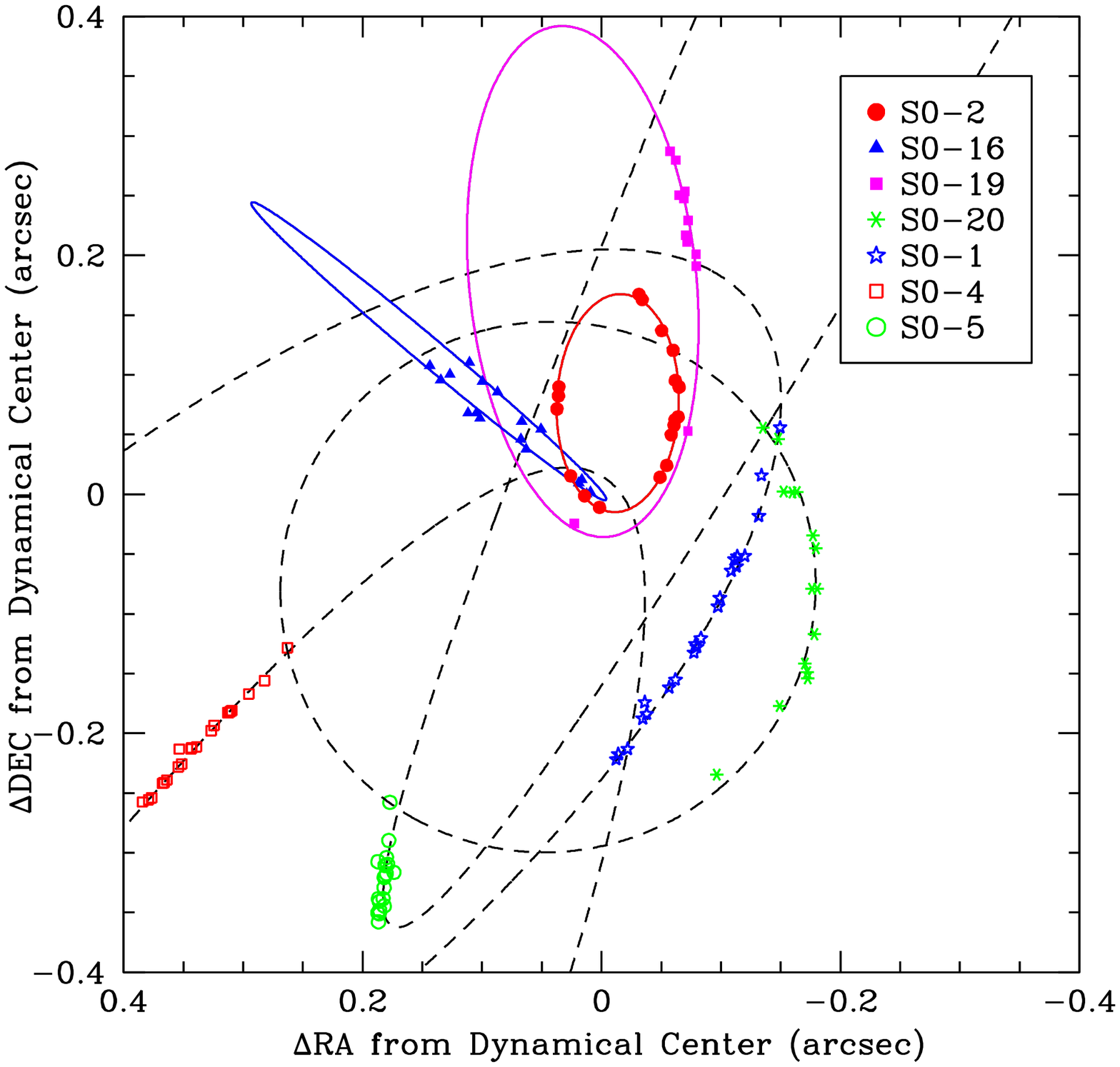}
\hfill
\includegraphics[height=2.65in]{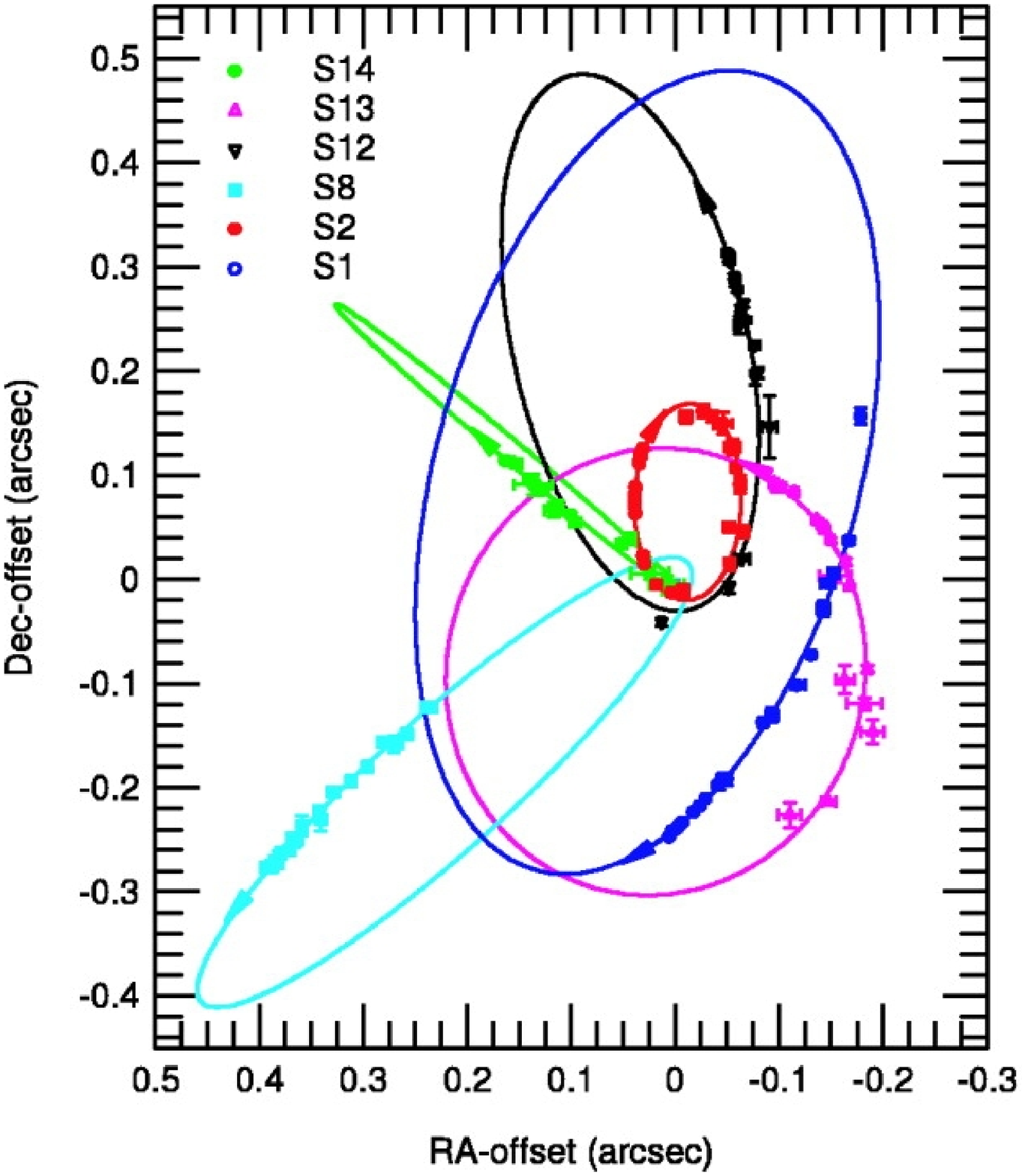}
\caption{Positions on the sky as 
a function of time for the central stars
orbiting the compact radio source Sgr~A*. Left: the data from the UCLA
group working with the Keck telescope (Ghez {\etal} \cite{ghe05b}). Right:
the data from the MPE--Cologne group at the ESO-VLT (Sch\"odel {\etal}
\cite{sch03}; Eisenhauer {\etal} \cite{eis05a}; Gillessen {\etal}, 
in preparation).
}\label{fig2}
\end{figure*}

\section{Compelling evidence for a central massive black hole}
With diffraction limited imagery starting in 1991 on the $3.5$m ESO New
Technology Telescope and continuing since 2002 on the VLT, a group at
MPE was able to determine proper motions of stars as close as
$\sim0.1$~arcsec from Sgr~A* (Eckart \& Genzel \cite{eck96},
\cite{eck97}). In 1995 a group at the University of California, Los
Angeles started a similar program with the $10$m diameter Keck telescope
(Ghez {\etal} \cite{ghe98}). Both groups independently found that the stellar
velocities follow Kepler laws and exceed $10^3$~km/s within the central
light month.

Only a few years later both groups achieved the next and crucial step:
they were able to determine individual stellar orbits for several stars
very close to the compact radio source (Fig.~\ref{fig2}; Sch\"odel et
al. \cite{sch02}, \cite{sch03}; Ghez {\etal} \cite{ghe03}, \cite{ghe05b};
Eisenhauer {\etal} \cite{eis05a}). In addition to the astrometric imaging
they obtained near-diffraction limited Doppler spectroscopy of the same
stars (Ghez {\etal} \cite{ghe03}; Eisenhauer {\etal} \cite{eis03a},b),
yielding precision measurements of the three dimensional structure of
several orbits, as well as the distance to the Galactic Center. At the
time of writing, the orbits have been determined for about a dozen stars
in the central light month. The central mass and stellar orbital
parameters derived by the two teams agree mostly very well. The orbits
show that the gravitational potential indeed is that of a point mass
centered on Sgr~A* within the relative astrometric uncertainties of
$\sim10$~milliarcsec. Most of the mass must be concentrated well within
the peri-approaches of the innermost stars, $\sim10$--$20$ light hours,
or $70$ times the Earth orbit radius and about $1000$ times the event
horizon of a $3.6$ million solar mass black hole. There is presently no
indication for an extended mass greater than about $5$\% of the point
mass.

Simulations indicate that current measurement accuracies are sufficient
to reveal the first and second order effects of Special and General
Relativity in a few years time (Zucker {\etal} \cite{zuc06}).
Observations with future $30$m+ diameter telescopes will be able to
measure the mass and distance to the Galactic Center to $\sim0.1$\%
precision. They should detect radial precession of stellar orbits due to
General Relativity and constrain the extended mass to $<10^{-3}$ of the
massive black hole (Weinberg, Milosavljevic \& Ghez \cite{wei05}). At
that level a positive detection of a halo of stellar remnants (stellar
black holes and neutron stars) and perhaps dark matter would appear to
be likely. Future interferometric techniques will push capabilities yet
further.

Long-term VLBA observations have set $2\sigma$ upper limits of about
$20$~km/s and $2$~km/s (or $50$~micro-arcsec per year) to the motion of
Sgr~A* itself, along and perpendicular to the plane of the Milky Way,
respectively (Reid \& Brunthaler \cite{rei04}; see also Backer \& Sramek
\cite{bac99}). This precision measurement demonstrates very clearly that
the radio source itself must indeed be massive, with simulations
indicating a lower limit to the mass of Sgr~A* of $\sim10^5$ solar
masses. The intrinsic size of the radio source at millimeter wavelengths
is less than $5$ to $20$ times the event horizon diameter (Bower {\etal}
\cite{bow04}; Shen {\etal} \cite{she05}). Combining the radio size and
proper motion limit of Sgr~A* with the dynamical measurements of the
nearby orbiting stars leads to the conclusion that Sgr~A* can only be a
massive black hole, beyond any reasonable doubt. An astrophysical dark
cluster fulfilling the observational constraints would have a life-time
less than a few $10^4$ years and thus can be safely rejected, as can be
a possible fermion ball of hypothetical heavy neutrinos. In fact all
non-black hole configurations can be excluded by the available
measurements (Sch\"odel {\etal} \cite{sch03}; Ghez {\etal} \cite{ghe05b})
-- except for a hypothetical boson star and the gravastar
hypothesis, but it appears that the two mentioned alternatives have
difficulties of their own, and they are less likely and certainly much
less understood than black holes (e.g. Maoz \cite{mao98}; Miller {\etal}
\cite{mil98}).
We thus conclude that, under the assumption of the validity of General
Relativity, the Galactic Center provides the best quantitative evidence
for the actual existence of (massive) black holes that contemporary
astrophysics can offer.

\begin{figure*}
\begin{center}
\includegraphics[width=0.7\textwidth]{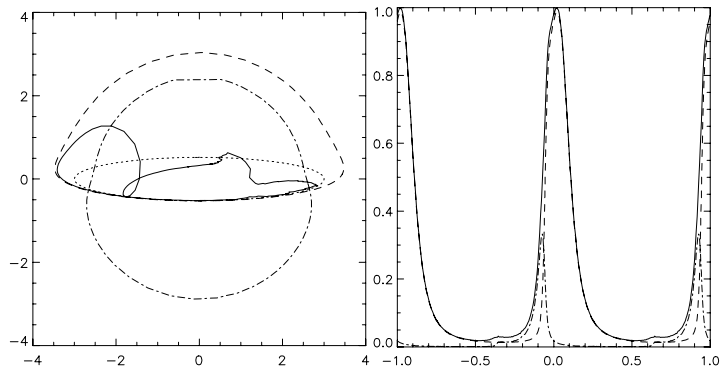}
\end{center}
\caption{Photo-center wobbling (left) and light curve (right) of a hot
spot on the innermost stable orbit around Schwarzschild black hole 
(inclination of $80$~deg), as derived from ray-tracing computations. 
Dotted curve: `true' path of the hot spot; dashed curves: apparent path
and a predicted light curve of the primary image; dash-dotted curves:
the same for secondary image; solid curves: path of centroid and
integrated light curve. Axes on the left panel are in Schwarzschild
radii of a $3$~million solar-mass black hole, roughly equal to the
astrometric accuracy of $10$~arcsec; the abscissa axis of the right
panel is in cycles. The loop in the centroids track is due to the
secondary image, which is strongly sensitive to the space-time
curvature. The overall motion can be detected at good significance at
the anticipated accuracy of GRAVITY. Details can be obtained by
analyzing several flares simultaneously (Gillessen {\etal} \cite{gil06b};
Paumard {\etal} \cite{pau05}).}\label{fig3}
\end{figure*}

\section{Zooming in on the accretion zone and event horizon}
Recent millimeter, infrared and X-ray observations have detected
irregular, and sometimes intense outbursts of emission from Sgr~A*
lasting anywhere between 30 minutes and a number of hours and occurring
at least once per day (Baganoff {\etal} \cite{bag01}; Genzel {\etal}
\cite{gen03a}; Marrone {\etal} \cite{mar06}). These flares originate from
within a few milli-arcseconds of the radio position of Sgr~A*. They
probably occur when relativistic electrons in the innermost accretion
zone of the black hole are significantly accelerated, so that they are
able to produce infrared synchrotron emission and X-ray synchrotron or
inverse Compton radiation (Markoff {\etal} \cite{mar01}; Yuan {\etal}
\cite{yua03}; Liu {\etal} \cite{liu05}). This interpretation is also
supported by the detection of significant polarization of the infrared
flares (Eckart {\etal} \cite{eck06b}), by the simultaneous occurrence of
X- and IR-flaring activity (Eckart {\etal} \cite{eck06a}; Yusef-Zadeh et
al. \cite{yus06}) and by variability in the infrared spectral properties
(Ghez {\etal} \cite{ghe05b}; Gillessen {\etal} \cite{gil06a}; Krabbe et
al. \cite{kra06}). There are indications for quasi-periodicities in the
light curves of some of these flares, perhaps due to orbital motion of
hot gas spots near the last circular orbit around the event horizon
(Genzel {\etal} \cite{gen03a}; Aschenbach {\etal}
\cite{asc04}; B\'elanger {\etal} \cite{bel06}). 

The infrared flares as well as the steady microwave emission from
Sgr~A* may be important probes of the gas dynamics and space-time metric
around the black hole (Broderick \& Loeb \cite{bro06}; Meyer {\etal}
\cite{mey06a},b; Paumard {\etal} \cite{pau06b}). Future long-baseline
interferometry at short millimeter or sub-millimeter wavelengths may be
able to map out the strong light-bending effects around the photon orbit
of the black hole. It is interesting to realize that the angular size of
the ``shadow'' of black hole (Bardeen \cite{bar73}) is not very far from
the anticipated resolution of interferometric techniques and it may thus
be accessible to observations in near future (Falcke, Melia \& Agol
\cite{fal00}).

Polarization measurements will help us to set further constraints on the
emission processes responsible for the flares. Especially the
time-resolved lightcurves of the polarized signal carry specific
information about the interplay between the gravitational and magnetic
fields near Sgr~A* horizon, because the propagation of the polarization
vector is sensitive to the presence and properties of these fields along
the light trajectories (Bromley, Melia \& Liu \cite{bro01}; Hor\'ak \&
Karas \cite{hor06}; Paumard {\etal} \cite{pau06b}). Polarization is also
very sensitive also to intrinsic properties of the source -- its
geometry and details of radiation mechanisms responsible for the
emission.

Synthesis of different techniques will be a promising way for the
future: the astrometry of central stars gives very robust results
because the stellar motion is almost unaffected by poorly known
processes of non-gravitational origin, while the flaring gas occurs much
closer to the black hole horizon and hence it directly probes the
innermost regions of Sgr~A*. Eventually the two components -- gas and
stars  of the Galaxy Center -- are interconnected and form the unique
environment in which the flaring gas is influenced by intense stellar
winds whereas the long-term motion and the `non-standard' evolution of
the central stars bears imprints of the gaseous medium though which the
stars pass.

Eisenhauer {\etal} (\cite{eis05b}) are developing GRAVITY (an instrument for 
`General Relativity Analysis via VLT Interferometry'), which will provide 
dual-beam, $10$ micro-arcsecond precision infrared astrometric imaging of faint 
sources. GRAVITY may be able to map out the motion on the sky of hot spots during
flares with a high enough resolution and precision to determine the size
of the emission region and possibly detect the imprint of multiple
gravitational images (see Fig.~\ref{fig3}). In addition to studies of
the flares, it will also be able to image the orbits of stars very close
to the black hole, which should then exhibit the orbital radial oscillations 
and Lense-Thirring precession due to General Relativity. Both the microwave
shadows as well as the infrared hot spots are sensitive to the
space-time metric in the strong gravity regime. As such, these
ambitious future experiments can potentially test the validity of the
black hole model near the event horizon and perhaps even the validity of
General Relativity in the strong field limit.

\begin{acknowledgments}
An extended version of this lecture was presented by RG as Invited Discourse 
during the $26^{\rm{}th}$~General Assembly of the International Astronomical Union
in Prague, 22nd August 2006 ({\em{}Highlights of Astronomy}, Volume~14, 2007). 
VK thanks the Czech Science Foundation for continued support (ref.\ 205/07/0052).
\end{acknowledgments}

\end{document}